\documentclass[a4paper,11pt]{article}
\usepackage{pos}

\newcommand{\beq}{\begin{equation}}
\newcommand{\eeq}{\end{equation}}
\newcommand{\bea}{\begin{eqnarray}}
\newcommand{\eea}{\end{eqnarray}}

\begin{document}

\title{On turbulent viscosity in relativistic jets and accretion disks}
\ShortTitle{On turbulent viscosity}
\author*[]{Alexander A. Panferov}
\affiliation[a]{Togliatty state university,\\Belorusskaya 14, Togliatty, Russia}
\emailAdd{panfS@yandex.ru}

\abstract{
The mechanism of turbulent viscosity is the central question in investigations of turbulence. This is also the case in the accretion disk theory, where turbulence is considered to be responsible for the outward transport of angular momentum in the accretion disk. In turbulent flows, vortices transport momentum over their length scales providing the mechanism of viscosity that is controlled by mass entrainment. We have earlier proposed an entrainment model for the particular case of the relativistic jets in the radio galaxy 3C\,31. In this paper, we further constrain the model parameters. The model (in the non-relativistic part) is successfully tested versus experimental and simulation data on the Reynolds stresses of free mixing layers and predicts the Smagorinsky constant $C_\mathrm{S} \approx 0.11$, which is consistent with the experimental range for shear flows $C_\mathrm{S} \approx 0.1-0.12$. For accretion disks, the entrainment model allows us to derive the same accretion mass rate as in the Shakura \& Sunyaev's $\alpha$-model without appealing to the turbulent kinematic viscosity $\nu_\mathrm{t}$, and the viscosity parameter $\alpha$ derived in the form $\displaystyle \alpha = -\frac{8}{3} \beta s_\mathrm{T} \frac{\mathrm{v_t}^2}{c_\mathrm{s}^2}$ depends on the power $s_\mathrm{T}$ of the temperature slope along the disk radius, $T\propto r^{s_\mathrm{T}}$, and quadratically on the turbulent velocity $\mathrm{v_t}$.
}
\FullConference{%
  The Multifaceted Universe: Theory and Observations -  2022 (MUTO2022)\\
  23-27 May 2022\\
  SAO RAS, Nizhny Arkhyz, Russia\\}

\maketitle

\section{Introduction }
\label{sec1}
In free shear flows, which jets and accretion disks look like, turbulent mixing layers develop and entrain matter from the surrounding irrotational flow by means of large-scale vortices \cite[Brown \& Roshko 2012]{BR12}. These vortices determine transverse transport of momentum, i.e. turbulent stress.  

The turbulent stress $\tau$ is usually dubbed as the Reynolds stress ($R$), which is determined by turbulent pulsations of fluid velocity, e.g. $\tau_\mathrm{xy} = -\rho \overline{\mathrm{v_x v_y}}$ for the (xy)-component of the turbulent stress tensor, which is supplementary to the stress tensor of the Reynolds equations system, where $\rho$ is the fluid density, $\mathrm{v_x}$ and $\mathrm{v_y}$ are the x and y-components of the fluid fluctuating velocity, respectively, thus $\mathrm{V_x} = \overline{\mathrm{V_x}} + \mathrm{v_x}$ is the whole x-component of the velocity. The Reynolds stresses cannot be derived from the system of Reynolds equations because the latter is unclosed. The usual approach to the turbulent stress is local, after Boussinesq, through the kinematic turbulent (eddy) viscosity coefficient $\nu_\mathrm{t}$ and local mean strain: $\displaystyle
\tau_\mathrm{xy} = - \rho \nu_\mathrm{t} \frac{\partial \overline{\mathrm{V_x}}}{\partial y}$
in the {\em simple} shear, where the gradient $\displaystyle \frac{\partial \overline{\mathrm{V_x}}}{\partial y}$ is only non-zero. The turbulent  viscosity may be modelled as 
$\displaystyle \nu_\mathrm{t} = \mathrm{v_t} l_\mathrm{m} = l_\mathrm{m}^2 \left| \frac{\partial \overline{\mathrm{V_x}}}{\partial y}\right| $
using the concept of Prandtl's mixing length $l_\mathrm{m}$, an empirical parameter. In this way Shakura \& Sunyaev \cite[thereafter SS73]{SS73} prescribed the turbulent fluid viscosity as $\nu_\mathrm{t} = \mathrm{v_t} l_\mathrm{m} = \alpha c_\mathrm{s} H$ in the $\alpha$-model of accretion disk, where $\displaystyle \alpha = \frac{\mathrm{v_t} l_\mathrm{m}}{c_\mathrm{s} H}$ is the viscosity parameter, $\mathrm{v_t}$ the turbulent velocity, $c_\mathrm{s}$ the speed of sound, $H$ the half thickness of the disk. However, the concept of turbulent viscosity does not represent the mechanism of turbulent stress, when large-scale vortices dominate the momentum transport which determines the stress \cite[Brown \& Roshko 2012]{BR12}.

In our paper \cite[Panferov 2017]{P17} we have derived a formula for the entrainment in the jets of  the radio galaxy 3C\,31 (repeated here in Section~\ref{sec2}). Here, we express the turbulent shear stress in dependency on turbulence intensity, elaborating the model of  Panferov \cite{P17}, and test this model vs. laboratory and simulation data (Section~\ref{sec3}). The model principles are successfully applied to explain the experimental value of the Smagorinsky constant, widely used in large eddy simulations of turbulent flows. After this verification we search for entrainment in accretion disks (Section~\ref{sec4}). Section~\ref{sec5} is concluding.

\section{\large Entrainment in jets of 3C\,31}
\label{sec2}
\begin{figure}[t]
\centering
\includegraphics[width=9cm]{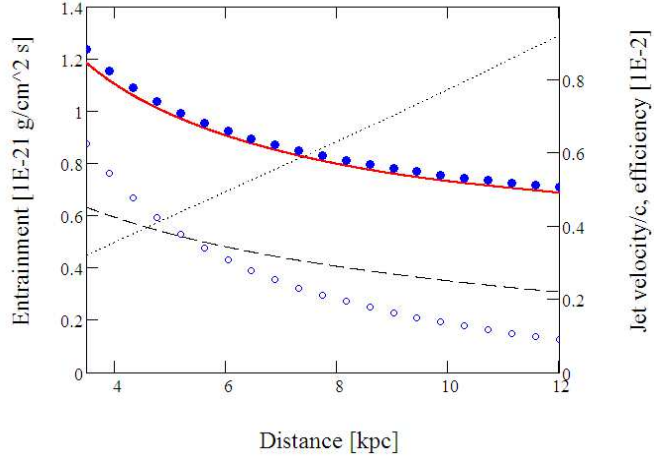}
\caption{\small
Profiles of the entrainment along the jets of 3C\,31: for the flux densities of the external (blue solid circles) and of the internal (blue hollow circles) entrainment after Wang et al. \cite{W09}, and the theoretical external entrainment $q_\mathrm{t}$ (red solid line). The internal entrainment is from Fig.~5a of Wang et al. \cite{W09}, not their approximation $g_\mathrm{s}(x)$. The jet velocity (dashed line) and the turbulence cross-section (dotted line) are also plotted.
}
\label{entr_prof}
\end{figure}

Jets of FR~I radio galaxies strongly decelerate at kiloparsec scales \cite[Laing and Bridle 2014]{LB14}. The origin of this is seemingly the turbulent friction of the jets in the ambient medium, which is associated with entrainment and is indicated by a shear at the edges in the transverse velocity profiles of the jets. The profile of the external entrainment along the jet 3C\,31 from Wang et al. \cite{W09}, which is shown in Fig.~\ref{entr_prof}, we have described using the function of mass flux density at the jet edge
\beq
q_\mathrm{t} = \beta \rho_\mathrm{a} \mathrm{v_t} \sigma_\mathrm{t},
\label{q}
\eeq
in dependency on the ambient medium density $\rho_\mathrm{a}$, the statistical coefficient $\beta$, the mean turbulent velocity $\mathrm{v_t}$ and the so-called cross-section of turbulence $\sigma_\mathrm{t}$ \cite[Panferov 2017]{P17}. In this formula, $\beta \rho_\mathrm{am} \mathrm{v_t}$ terms the mean flux density of mass in any direction driven by the turbulence at the jet edge. And the cross-section 
\beq
\sigma_\mathrm{t} = \sin i = \frac{\mathrm{v}_\perp} {\sqrt{\mathrm{v}_\parallel^2+\mathrm{v\mathrm{_\perp^2}}}} = \frac{\eta}{\sqrt{\gamma^2+\eta^2}} 
\label{sig_t}
\eeq
is the efficiency of the capture of matter from vortices, which are imagined as fluid pulsations, by the vortical turbulent flow of the jet of the velocity $\mathrm{v_j}$. Here $\mathrm{v}_\parallel = \mathrm{v_j}$, $\displaystyle \mathrm{v}_\perp = \frac{\mathrm{v_{t\perp}}}{\gamma}$ are the parallel and perpendicular components of the pulsation velocity (e.g. $\mathrm{v_{t\perp}}^2 = \mathrm{v_y}^2 + \mathrm{v_z}^2$), $i$ the inclination to the jet of the turbulence pulses, in the jet rest frame, $\displaystyle \eta= \frac{\mathrm{v_{t\perp}}}{\mathrm{v_j}}$, $\displaystyle \gamma = \left(1-\left(\frac{\mathrm{v_j}}{c}\right)^2\right)^{-1/2}$ the Lorentz factor, $c$ the speed of light. 

%
%

Two parameters of the entrainment model, $\beta$ and $\mathrm{v_t}$, were defined intuitively as $\displaystyle \frac{1}{6}$ and the speed of sound $c_\mathrm{s}$, respectively, and the entrained matter was supposed to just acquire the jet velocity $\mathrm{v_j}$. The latter supposition straightforwardly gives the turbulent shear stress at the jet edge $\tau_\mathrm{t} = q_\mathrm{t} \mathrm{v_j}$, or $\tau_\mathrm{t} = \alpha P$, where $\alpha$ now is the parameter dependent on turbulence intensity, not on turbulent viscosity, and $P$ the pressure. Afterwards, $\beta$ was constrained to $\displaystyle \frac{1}{5}$ by fitting the function $q_\mathrm{t}$, which is plotted in Fig.~\ref{entr_prof}, to external entrainment of Wang et al. \cite{W09}. The entrainment model is simple and was compared with very speculative data, therefore its success on this plot seems as miraculous, and further investigations of the model are required.

\section{\large Parameters of entrainment model}
\label{sec3}
For isotropic turbulence the mean velocity of stochastic flow in a direction is, as follows:
\beq
\mathrm{v_f} = \beta \mathrm{v_t} = \mathrm{v_t} \int_0^{\pi/2} \cos{\theta} \frac{2\pi \sin{\theta} d\theta}{4\pi} = \frac{1}{4} \mathrm{v_t} = \frac{1}{4} \sqrt{\frac{3}{2}} \mathrm{v}_\perp,
\label{v_iso}
\eeq
where $\theta$ is the inclination of the velocity of a pulsation to the given direction. Then Eq.~(\ref{q}) is rewritten as
\beq
q_\mathrm{t} = \beta \rho_\mathrm{a} \mathrm{v}_\perp \sigma_\mathrm{t},
\label{q_m}
\eeq
where $ \displaystyle \beta = \frac{1}{3.266} \approx \frac{1}{\pi}$ is accepted, and $\mathrm{v}_\perp$ substitutes $\mathrm{v_t}$.

To test this model vs. experimental data on free shear flows, we have to invent how the entrainment $q_\mathrm{t}$ depends on turbulent stresses. For this we define the drag velocity $\mathrm{v}_\tau$ so that 
\beq
\tau_\mathrm{xy} = q_\mathrm{t} \mathrm{v}_\tau,
\label{stress_q}
\eeq
and the mean flux of the fluctuating streamwise momentum $\rho \mathrm{v_x}$, which is transported by the transverse fluctuations $\mathrm{v_y}$, or the turbulent shear stress, is derived for a self-similar mixing layer as follows:
\beq
\tau_\mathrm{xy} = q_\mathrm{t} \int_{-\infty}^{y_0} \rho \mathrm{V_x}(y)^2 dy\bigg/ \int_{-\infty}^{y_0} \rho \mathrm{V_x}(y) dy,
\label{v_drag}
\eeq
where $y_0$ is the depth where $\tau_\mathrm{xy}(y_0)$ is maximal, from which we can express $\mathrm{v}_\tau$. Hereafter the term $\mathrm{V_x}$ means the streamwise velocity averaged over time. Hence, the drag velocity is the weighted stream velocity acquired by entrained matter, it determines the stress, therefore it substitutes the variable $\mathrm{v_j}$ in the turbulent cross-section (\ref{sig_t}), as we guess.

%
\begin{figure}[h]
\centering
\includegraphics[width=9cm]{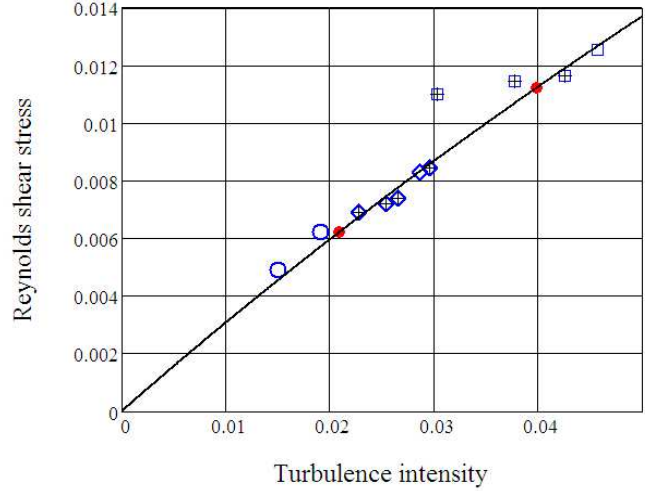}
\caption{\small 
A graph of the experimental (red full circles) and simulated (blue hollow symbols) data on the Reynolds stresses of self-similar free mixing layers in the coordinates of the squared intensity $\displaystyle \frac{R_\mathrm{yy} + R_\mathrm{zz}}{\rho \mathrm{V_x}^2}$ and the normalized shear stress $\displaystyle \frac{R_\mathrm{xy}}{\rho \mathrm{V_x}^2}$. The experimentals are from \cite[Bell \& Mehta 1990]{BM90} and \cite[Gruber et al. 1993]{G93},the data of spatial direct numerical simulations (DNS) are from \cite[Zhou et al. 2012]{Z12} and \cite[Barre \& Bonnet 2015]{BB15} (circles), the data of variable-density temporal DNS are from \cite[Almagro et al. 2017]{A17} and \cite[Baltzer \& Livescu 2020]{BL20} (squares and diamonds, respectively), crosses stand for the density ratio of the two free streams $s\neq 1$. Line is the function $\tau_\mathrm{xy} = q_\mathrm{t} \mathrm{v}_\tau$.  
}
\label{labs}
\end{figure}
%

We have collected the experimental and simulated data on {\em self-similar} free mixing layers from the literature. To the best of our knowledge, there are only few experimental data on the needed set of Reynolds stresses ($R_\mathrm{xy}$, $R_\mathrm{yy}$ and $R_\mathrm{zz}$): the data on $R_\mathrm{zz}$ are very scarce, besides they are contradictory to each other \cite[Zhou et al. 2012]{Z12}. The simulated data on spatial free shear flows also are scarce, and the temporal ones are not quite adequate for the entrainment investigations. The collected data are plotted in Fig.~\ref{labs} in coordinates of the normalized Reynolds stresses: $\displaystyle (I_\mathrm{y}^2 + I_\mathrm{z}^2,\ I_\mathrm{xy}^2) \equiv \frac{(R_\mathrm{yy} + R_\mathrm{zz},\ R_\mathrm{xy})}{\rho \mathrm{V_x}^2}$, where $\mathrm{V_x}$ is the velocity difference over the shear layer width. 

The plot of Eq.~(\ref{stress_q}) is well aligned with the plotted data points and exactly predicts the point of Bell \& Mehta. This is the only point among ocean of experimental data on turbulent mixing layers which is characterized by uncertainty, and this uncertainty is only $\pm 5\%$, while the observed experimental scatter is 25-50\% for the layer growth rates (and the same is true for the simulations, \cite[Smits \& Dussauge 2006, pp. 140, 156]{SD06}). Moreover, this point is the most referenced in comparative studies. Thus, other points have much smaller weights and we don't discuss them.

To take a more hard probe of the entrainment law (\ref{q_m}), we checked out its capability to predict the Smagorinsky constant $C_\mathrm{S}$, an empirical constant used in turbulent flow models in large eddy simulations (e.g. \cite[Pope 2000, p. 587]{P00}). For the {\em simple} shear layer we match the kinematic turbulent viscosity after  Smagorinsky to the viscosity trivially derived from Eq.~(\ref{stress_q}) omitting the square root in $\sigma_\mathrm{t}$:
\beq
\nu_\mathrm{t} = (C_\mathrm{S} \Delta_\mathrm{S})^2 \left| \frac{\mathrm{d V_x}}{\mathrm{d}y} \right| \equiv \beta \mathrm{v}_\perp^2 \left(\frac{\mathrm{d}y}{\mathrm{d V_x}}\right)^2 \left| \frac{\mathrm{d V_x}}{\mathrm{d}y} \right| \approx \beta \left(\frac{\mathrm{v}_\perp}{\Delta \mathrm{V_x}} \Delta y\right)^2 \left| \frac{\mathrm{d V_x}}{\mathrm{d}y} \right|,
\label{visc_Smag}
\eeq
where $\Delta_\mathrm{S}$ is the specific length scale of large eddy simulations, and the mean gradient over the transverse scale $\Delta y$ substitutes the local gradient. By setting $\Delta y \equiv \Delta_\mathrm{S}$ and substituting the experimental turbulent intensities of self-similar incompressible plane mixing layers from \cite[Yoder et al. 2015]{Y15}
\beq
0.016 \le \frac{\overline{\mathrm{v_y}^2}}{\Delta \mathrm{V_x}^2} \le 0.020,\
0.020 \le \frac{\overline{\mathrm{v_z}^2}}{\Delta \mathrm{V_x}^2} \le 0.022,
\label{I_Y}
\eeq
for the ratio $\displaystyle \bigg(\frac{\mathrm{v}_\perp}{\Delta \mathrm{V_x}}\bigg)^2$ in Eq.~(\ref{visc_Smag}) we get the constant value range:
\beq
C_\mathrm{S} = \sqrt{\beta} \frac{\mathrm{v}_\perp}{\Delta \mathrm{V_x}} = 0.107 - 0.116,
\label{C_S}
\eeq
which is in the bounds of the experimental range $C_\mathrm{S} \simeq 0.1 - 0.12$ for shear flows \cite[Sagaut 2006, p. 124]{Sag06}. Omitting the square root in $\sigma_\mathrm{t}$ we exaggerated the above $C_\mathrm{S}$ by $\approx 7\%$, if to take roughly $\displaystyle \mathrm{v}_\tau = \frac{\mathrm{V_x}}{2}$, that does not change positivity of our test of the entrainment law vs. the Smagorinsky constant. This justifies the setting $\Delta y \equiv \Delta_\mathrm{S}$ that, on other hand, suggested proportionality of the scales of turbulent velocity $\mathrm{v_t}$ to the driving scales $\Delta \mathrm{V_x}$, i.e. $\displaystyle \frac{\mathrm{v}_\perp}{\Delta \mathrm{V_x}}$ is independent on $\Delta y$.

\section{\large Entrainment in accretion disk}
\label{sec4}
An accretion disk is a rotating shear layer. Nonlaminar flows are inevitable in accretion disks in view of their huge Reynolds numbers. Pure hydrodynamic turbulence and the formation of even large long-living vortices in accretion disk models has been observed \cite[Manger \& Klahr 2018]{Man18}. Vortices, we know, provide turbulent shear stress over macroscopic distances of mixing. We conjecture that long-living shear layers of intermittent vorticity, irrespectively of  mechanisms of turbulence, form in the accretion disks (see vorticity maps in \cite[Manger \& Klahr 2018]{Man18} and \cite[Richard, Nelson \& Umurhan 2016]{Rich16}) and entrain the disk stuff from the irrotational layers between them, i.e. pull the stuff radially inward and outward of the disk, with the net effect of stuff accretion and angular momentum excretion in the entrainment processes. 

In a steady mode of accretion with the accretion rate of mass inflow $\dot M_\mathrm{a}$, the accretion disk between radii $r$ and $r_0$ has to shed the angular momentum
\beq
\dot M_\mathrm{a} (\mathrm{v}_\phi r - \mathrm{v}_{\phi 0} r_0) =  T_\mathrm{r\phi} r - T_\mathrm{r\phi 0} r_0,
\label{ang}
\eeq
due to the momentum fluxes $T_\mathrm{r\phi}$. Here $\dot M_\mathrm{a}$ and $T_\mathrm{r\phi}$ (and the entrainment flux $Q_\mathrm{t}$ below) are the integrated rates over a cylindrical cross-section of the accretion disk of a radius $r$, e.g. $\displaystyle T_\mathrm{r\phi} = \int_{-\infty}^{\infty}\int_{0}^{2\pi} \tau_\mathrm{r\phi} dzd\phi$ in cylindrical coordinates ($r,\,\phi,\,z$), and $\mathrm{v}_\phi$ is the azimuthal velocity. Under the condition $T_\mathrm{r\phi 0}=0$, the momentum flux
\beq
T_\mathrm{r\phi} = \dot M_\mathrm{a} \mathrm{v}_\phi \left(1 - \left(\frac{r}{r_0}\right)^{1/2}\right),
\label{ang_mom}
\eeq
which is resolved from Eq.~(\ref{ang}) (cf. Eq.~(2.4) in \cite[SS73]{SS73}, here $\mathrm{v}_\phi$ is allowed to be the Keplerian), can be induced:\\
a) either by "the friction between the adjacent layers" \cite[SS73]{SS73}, i.e. locally; in this way $\dot M_\mathrm{a}$ is derived by Shakura \& Sunyaev \cite[SS73]{SS73} as $\dot M_\mathrm{a} = f(T_\mathrm{r\phi})$;\\
b) or by large-scale vortices, i.e. by entrainment; then $\dot M_\mathrm{a} = \Delta Q_\mathrm{t} = Q_\mathrm{t\,in} - Q_\mathrm{t\,out}$, in this way the accretion rate is not a function of turbulent viscosity.\\

The entrainment in the disk is locally seen as that in a flat mixing layer, for which the formula (\ref{q_m}) was designed, if the disk half thickness $H \ll r$. For convenience, we again use it in the stripped down version:
\beq
q_\mathrm{t} = \beta \rho \frac{\mathrm{v}_\perp^2}{\mathrm{v}_\tau}.
\label{q_v2}
\eeq
As in the case discussed above, the fluctuations $\mathrm{v}_\perp$ are proportional to the driving scale $\mathrm{v}_\tau$, which, again, roughly equals a half of the (Keplerian) velocity difference in the ring turbulent layer of a width $\Delta r$. The disk half thickness $H$ is defined from the disk hydrostatic balance as $\displaystyle \frac{H}{r}=\frac{c_\mathrm{s}}{\mathrm{v}_\phi}$, therefore the drag velocity, and the turbulent velocity $\mathrm{v}_\perp$ too, are scaled with the temperature of the disk as
\beq
\mathrm{v}_\tau = \frac{1}{2}\left|\Delta \mathrm{v}_\phi\right| \approx \frac{1}{4} \frac{\mathrm{v}_\phi}{r} \Delta r = \frac{1}{4} \frac{\Delta r}{H} c_\mathrm{s} \propto T^{1/2}.
\label{v_drag_T}
\eeq
In accretion disks, temperature slopes as $\displaystyle T \propto r^{s_\mathrm{T}}$, where the power $s_\mathrm{T}<0$. The entrainment fluxes $Q_\mathrm{t\,in}$ and $Q_\mathrm{t\,out}$ are the time-averages over the same surface but of opposing vortical flows. Here we conjecture that the mean density $\rho$ in the opposing vortices does not change essentially while they propagate, and their stochastic velocities adjust to the temperature of the ambient medium (really, to the driving velocity, which is the property of the background flow). Thus we get the entrainment scaling only with $r$, $\displaystyle q_\mathrm{t} \propto r^{s_\mathrm{T}/2}$. Therefore, the accretion rate is derived by a simple difference of the entrainment fluxes $Q_\mathrm{t}$ of the neighboring mixing layers, each of the same width $\Delta r \ll r$:
\beq
\dot M_\mathrm{a} = 2\pi r \Sigma \beta \left(\frac{\mathrm{v}_\perp^2}{\mathrm{v}_\tau}\bigg|_{r-\Delta r} - \frac{\mathrm{v}_\perp^2}{\mathrm{v}_\tau}\bigg|_{r+\Delta r}\right) \approx - 2\pi \Sigma \beta \frac{s_\mathrm{T}}{2} \frac{\mathrm{v}_\perp^2}{\mathrm{v}_\tau} 2\Delta r,
\label{M_a}
\eeq
where $\displaystyle \Sigma = \int_{-\infty}^{\infty}\rho dz$ is the surface density in the disk. Substituting expression (\ref{v_drag_T}) for $\mathrm{v}_\tau$ in this equation, we arrive to:
\beq
\dot M_\mathrm{a} \omega= 2\pi \alpha \Sigma c_\mathrm{s}^2,
\label{Ma_SS}
\eeq
where $\omega$ is the angular velocity, and
\beq
\alpha = -\frac{8}{3} \beta s_\mathrm{T} \frac{\mathrm{v_t}^2}{c_\mathrm{s}^2}.
\label{alpha_E}
\eeq
The functional form of the accretion rate in Eq.~(\ref{Ma_SS}) is just the same as in  the $\alpha$-model of accretion disks of Shakura \& Sunyaev \cite[SS73, Eq.~(2.4)]{SS73} for $r \gg r_0$. Moreover, the viscosity parameter $\alpha$ has the right functional form $\displaystyle \alpha \propto \frac{\mathrm{v_t}^2}{c_\mathrm{s}^2}$, which can be envisaged from the physical interpretation of the stress $\displaystyle \tau_\mathrm{r\phi} \propto \rho \mathrm{v_t}^2$ and the definition $\displaystyle \tau_\mathrm{r\phi} \propto \alpha \rho c_\mathrm{s}^2$ of the parameter $\alpha$ \cite[SS73, Eq.~(1.2)]{SS73}. 

\section{\large Conclusions}
\label{sec5}
We have elucidated the entrainment model of \cite[Panferov 2017]{P17} and applied it to the problems on turbulent shear stress, in other words, on turbulent viscosity, in a plane mixing layer, in the jets of 3C\,31 and in accretion disks. The model reduces the problem of turbulent shear stress to the question about turbulence intensity. 

The entrainment model allows us to predict the deceleration of the 3C\,31 jets, which is caused by turbulent friction at the jets surface. The model readily expresses the surface turbulent stress in the form $\tau_\mathrm{t} = \alpha P$, where $\alpha$ can be set.

Interdependence of the Reynolds stresses observed in plane mixing layers under varying physical conditions is predicted (Fig.~\ref{labs}).

Application of the principles of the entrainment model to the Smagorinsky prescription of turbulent viscosity gave us the Smagorinsky constant $C_\mathrm{S} \simeq 0.11$, observed in experimental shear flows.

The entrainment model has been shown to be capable to construct the widely used $\alpha$-model of accretion disks in another way, bypassing the Prandtl's assignment of turbulent viscosity.

\section*{\large Acknowledgements}
I wish to thank the anonymous referee.

\end{document}